\documentstyle[12pt]{article}

\newcommand{\matel}[3]{\langle #1|#2|#3\rangle}
\newcommand{\ra}{\rightarrow}

\newcommand{\sG}{\sigma \cdot G}

\newcommand{\aver}[1]{\langle #1\rangle}

\newcommand{\as}{\alpha_s}

\newcommand{\MeV}{\,\mbox{MeV}}

\newlength{\dinwidth}
\newlength{\dinmargin}
\begin{document}

\vspace{1cm}

\begin{flushright}
UND-HEP-95-BIG06\\
June 1995\\
\end{flushright}

\begin{center}
\begin{Large}
\begin{bf}
\vspace{2cm}
%
%
LIFETIMES OF HEAVY-FLAVOUR HADRONS --
WHENCE AND WHITHER?
\footnote{Invited Lecture given at the 6th International
Symposium on Heavy Flavour Physics, Pisa, Italy,
June 6-10, 1995}
\\
\end{bf}
\end{Large}
\vspace{5mm}
\begin{large}
%
%
I.I. Bigi\\
\end{large}
%
%
Physics Dept., University of Notre Dame du Lac,
Notre Dame, IN 46556, U.S.A.\\
e-mail address: BIGI@UNDHEP.HEP.ND.EDU
\vspace{5mm}
\end{center}
\noindent
\begin{abstract}
A theoretical treatment for the
weak decays of
heavy-flavour hadrons has been developed
that is genuinely
based on QCD. Its methodology as it applies to
total lifetimes and the underlying theoretical issues are
sketched. Predictions are compared with present
data. One discrepancy emerges: the beauty baryon lifetime
appears to be significantly shorter than expected. The
ramifications of those findings are analyzed in detail.
\end{abstract}
%
{\bf I. Introduction: The Holy Grail and The Hope}

\noindent While the fertile minds of theorists have
spawned many
ideas formulated in the world of quarks and hadrons,
they have been conspicuously less productive in translating
these ideas into the world of hadrons.
Heavy-flavour physics presents us with
two
new perspectives onto this long-standing embarrassment
to the theoretical community.

\noindent (i) A detailed study of beauty
decays in particular addresses essential elements of the
Standard Model (hereafter referred to as {\em SM})
and thus provides us with fundamental
probes of it: what are the values of the KM parameters
$V(cb)$, $V(ub)$, $V(td)$ and $V(ts)$; can the {\em SM}
account for $B^0 - \bar B^0$ oscillations and rare $B$ decays;
finally, the `Holy Grail' of beauty physics: CP
violation. Hadronization should not be seen as exclusively
evil in that context. For without it particle-antiparticle
oscillations would not occur and it would become even moreJ
difficult for CP violation embedded in the quark Lagrangian
to manifest itself in observable transitions.

\noindent (ii) On the quark level there obviously exists
a {\em single} lifetime for a given flavour.
Differences in the lifetimes
of weakly decaying hadrons carrying the same flavour
thus provide us with a yardstick for evaluating the impact
of hadronization.  From $\tau (K^+)/\tau (K_S)\simeq 140$,
$\tau (D^+)/\tau (D^0)\simeq 2.5$ and
$0.9 \leq \tau (B^-)/\tau (B_d)\leq 1.2$ one infers that
deviations of the
lifetime ratios from unity decrease monotonically with the
heavy-flavour mass $m_Q$ increasing. This is as expected,
and it suggests that hadronization
effects can be addressed through an expansion in
$1/m_Q$. Heavy flavour decays thus constitute an
intriguing lab to study QCD in a novel environment with a
new probe, namely $m_Q$. This give us reason to hope
that nonperturbative effects can be brought under
theoretical control in beauty decays and maybe even
in charm decays.

There exist now four second-generation theoretical technologies
providing us with tools
to deal with heavy-flavour decays, namely QCD sum
rules; Monte Carlo simulations of QCD on the lattice; heavy quark
effective theory (HQET) and $1/m_Q$ expansions.
These technologies are genuinely based on QCD without
a need to invoke a
`deus ex machina' \footnote{It should be kept in mind,
though, that they all require an assumption concerning
quark-hadron duality that
has not been proven yet; I will briefly comment on that later on.}.
Among those only the $1/m_Q$ expansion allows
to treat inclusive decays. HQET, for example,
deals with formfactors in exclusive semileptonic transitions
rather than
total widths; those -- through the phase space --
depend strongly on
$m_Q$, whereas it is the special feature of HQET that $m_Q$ is
removed from its Lagrangian. However the
$1/m_Q$ expansion benefits from the
assistance of the other three techniques.

It is actually somewhat misleading to use the term
$1/m_Q$ expansion. For it is primarily the inverse of the
{\em energy release} rather than $1/m_Q$
that provides the expansion parameter.
The highest accuracy can thus be expected
for $b\ra u l \nu$ transitions, followed by
$b\ra cl \nu$ and $b\ra u \bar ud$ decays;
a lower precision holds
for $b\ra c \bar ud$ and $b\ra u \bar cs$;
the applicability of these tools to the $b\ra c \bar cs$
channels on the other hand is suspect --
a concern I will repeatedly return to.

The remainder of my talk will be organized as follows:
in Sect.II I will summarize the early phenomenology
before
introducing the $1/m_Q$ methodology for
fully integrated rates
in Sect.III; in Sect.IV I describe some applications and compare
the resulting predictions with present data before concluding in
Sect.V. In presenting the material I adopt three
guidelines, namely
to stress intuitive arguments over more formal ones (those
are given elsewhere); to place the arguments into today's
theoretical landscape and to emphasize the practical
relevance for experimental studies.  A more detailed discussion can
be found in Ref.\cite{REPORT}.

{\noindent
\bf II. Early Phenomenology: Myths, Legends and Truths}

\noindent Having been raised in Bavaria and being now in the
service of a catholic university, I appreciate that myths and
legends quite often contain more than just a kernel of truth.
However it typically is very difficult to ascertain
{\em beforehand} what is truth and what is poetic license (or worse).
The same situation applies with respect to phenomenological
models.

The starting point is the {\em spectator} process which
contributes {\em uniformly}
to the widths of all hadrons $H_Q$ of a given flavour. It rises so
rapidly with $m_Q$, namely
$ \Gamma _{spect}\propto G_F^2 m_Q^5$ (for $m_Q < M_W$)
that it dominates for large $m_Q$ and thus provides
the yardstick by which the non-leading (in $m_Q$)
contributions are evaluated. One mechanism was identified
for generating differences in the $H_Q$ lifetimes, namely
"Weak Annihilation" (=WA) of $Q$ with the light
valence antiquark for mesons or
"W Scattering" (=WS) with the valence diquark system for
baryons \footnote{A distinction is often made between W exchange
in the
s and in the t channel with the former case referred to
as `weak annihilation' and the latter as `W exchange'. This
classification
is however artificial since the two operators mix already under
one-loop renormalization in QCD.}.
Such an analysis had first been undertaken
for charm decays.
Since WA contributes to Cabibbo allowed decays of $D^0$, but not of
$D^+$
mesons (in the valence quark description), it creates a difference in
$\tau (D^0)$ vs. $\tau (D^+)$. However the WA rate to
lowest order in the strong interactions is doubly
suppressed relative to the spectator rate, namely by the helicity
factor $(m_q/m_c)^2$ with
$m_q$ denoting the largest mass in the final state and by the
`wavefunction overlap' factor $(f_D/m_c)^2$
reflecting the practically zero
range of the low-energy weak interactions:
$\Gamma ^{(0)}_{W-X}(D^0)
\propto G_F^2f^2_Dm_q^2m_c$.
Therefore it had
originally been suggested that already charm hadrons should
possess approximately equal lifetimes. It then came
as quite a
surprise when observations showed it to be otherwise -- in
particular since the first data suggested a considerably larger
value for $\tau (D^+)/\tau (D^0)$  than measured today. This
caused a re-appraisal of the theoretical situation; its
results {\em at that time} can be summarized in
three main points.

\noindent (i)  There is another significant source for
a lifetime difference
\cite{PI}. Cabibbo-allowed
nonleptonic $D^+$ decays
-- but not $D^0$ decays --
produce two antiquarks in the final state that
carry the same flavour:
$D^+ = [c\bar d] \ra (s\bar d u) \bar d$.
Thus one has to allow for the {\em interference}
between different
quark diagrams in $D^+$ , yet not in $D^0$ decays; the
$\bar d$ valence antiquark in $D^+$ mesons thus ceases to
play the role of an uninvolved bystander and a difference
in $\tau (D^+)$ vs. $\tau (D^0)$ will arise.  This interference
turns out to be destructive, i.e. it prolongs $\tau (D^+)$
over $\tau (D^0)$, but only once the QCD radiative corrections
have been included! This effect is usually referred to as
`Pauli Interference' (=PI) although such a name would be
misleading if it is interpreted as suggesting that
the interference
is automatically destructive.

\noindent (ii) It was argued \cite{SONI} that the
helicity suppression
of the WA contribution to $D$ decays can be vitiated. Evaluating
explicitely a
$W$-exchange diagram with gluon bremsstrahlung off the
initial antiquark line one finds:
$$\Gamma ^{(1)}_{W-X}(D^0)\propto
(\as /\pi ) G_F^2(f_D/\aver{E_{\bar q}})^2m_c^5
\eqno(1)$$
with
$\aver {E_{\bar q}}$ denoting the average energy of the
initial antiquark $\bar q$. Using a non-relativistic
wavefunction for the decaying meson one has
$\aver{E_{\bar q}}\simeq m_q$. This
contribution, although of higher order in $\as$, would
dominate over the lowest order term
$\Gamma ^{(0)}_{W-X}$ since helicity suppression has
apparently been vitiated and the decay constant $f_D$
is now
calibrated by $\aver{E_{\bar q}}$ with
$f_D/\aver{E_{\bar q}} \sim {\cal O}(1)$ rather than
$f_D/m_c \ll 1$.
The spectator picture would still apply at asymptotic quark
masses, since $\Gamma ^{(1)}_{W-X}/
\Gamma _{spec}\propto (f_D/\aver{E_{\bar q}})^2 \ra 0$
as $m_c\ra \infty$ due to $f_D \propto 1/\sqrt{m_c}$.
Eq.(1) -- if true -- would have a
dramatic impact on the theoretical description of weak
heavy-flavour decays: WA would be enhanced considerably and
be quite significant even in beauty decays.
Alternatively
it had been suggested \cite{MINKOWSKI}
that the wavefunction of the $D$ meson
contains a $c \bar q g$ component where the $c \bar q$ pair forms
a spin-one configuration with the gluon $g$ balancing the spin
of the $c\bar q$ pair.

\noindent Both effects listed above, namely PI and WA,
prolong $\tau (D^+)$ over $\tau (D^0)$.

\noindent (iii) A very rich structure emerges in the decays of
charm baryons \cite{BARYONS1}:
WS is {\em not} helicity suppressed
already to lowest order in the strong coupling;
PI affects the
$\Lambda _c$, $\Xi _c^{0,+}$ and $\Omega _c$
widths in various
ways, generating destructive as well as constructive contributions!
It is then very hard to make
reliable numerical predictions for these baryonic lifetimes
beyond
the overall qualitative pattern:
$$\tau (\Xi _c ^0) < \tau (\Lambda _c) < \tau (\Xi _c ^+) \eqno(2)$$

\noindent Reviews of these phenomenological descriptions
can be found in \cite{RUCKL}.

As it turned out, some
of the phenomenological descriptions anticipated the
correct results: it is PI that provides the main
engine behind
the $D^+$-$D^0$ lifetime ratio; $\Lambda _c$ is considerably
shorter-lived than $D^0$; the observed charm baryon
lifetimes do obey the hierarchy stated in eqs.(2).

Nevertheless the phenomenological treatments had significant
shortcomings, both of a theoretical and of a phenomenological
nature: (i) No agreement had emerged in the literature about how
corrections in particular due to WA and WS scale with the heavy
quark mass $m_Q$. (ii) Accordingly no clear predictions could be
made on the lifetime ratios among beauty hadrons, namely whether
$\tau (B^+)$ and $\tau (B_d)$ differ by a few to several percent
only, or by 20 - 30 \%, or by even more! (iii) No unequivocal
prediction on
$\tau (D_s)$ or $\tau (B_s)$ had appeared. (iv) In the absence of a
systematic
treatment it is easy to overlook relevant contributions, and that is
actually what happened; or the absence of certain corrections had to
be postulated in an ad-hoc fashion. Thus there existed both an
intellectual and a practical need for a description based on a
systematic theoretical framework rather than a set of phenomenological
prescriptions; this is provided by the $1/m_Q$
expansion.

\pagebreak
{\noindent
\bf III. Methodology of the Heavy Quark Expansion for Total
Rates}

\noindent To begin with, the decay dynamics have to be known on the
quark level. The charged current operators for a given
combination of quark
flavours are naturally defined at
scale $M_W$. The matrix elements for the decay process
are evaluated at an ordinary hadronic scale $\mu _{had}$;
therefore
one has to evolve these operators from $M_W$ down to
$\mu _{had}$, which is
done perturbatively. Since $\mu ^2_{had} \ll m^2_Q \ll M^2_W$
one finds that the perturbative QCD corrections have a very
sizeable impact on
decay rates. Yet they do not generate any lifetime
differences among the hadrons $H_Q$.

The size of the matrix elements
is controlled by nonperturbative
dynamics. It is here where the $1/m_Q$
expansion benefits from the results of the other three
technologies, since those can determine the size of some of
the relevant expectation values.

In analogy to the treatment of
$e^+e^-\rightarrow hadrons$ one describes the transition rate into an
inclusive final state $f$ through the imaginary part of a
forward scattering operator evaluated to second order in the weak
interactions \cite{SV,BUV}:
$$\hat T(Q\rightarrow f\rightarrow Q)=
i \, Im\, \int d^4x\{ {\cal L}_W(x){\cal L}_W^{\dagger}(0)\}
_T\eqno(3)$$
where $\{ .\} _T$ denotes the time ordered product and
${\cal L}_W$ the relevant effective weak Lagrangian expressed on
the
parton level. If the energy release in the decay is sufficiently large
one can express the {\em non-local} operator product in eq.(3) as an
infinite sum of {\em local} operators $O_i$ of increasing dimension
with
coefficients $\tilde c_i$
containing higher and higher inverse powers of
$m_Q$. The width for $H_Q\rightarrow f$ is then obtained by
taking the
expectation value of $\hat T$ between the state $H_Q$.
For semileptonic and nonleptonic
decays treated through order $1/m_Q^3$ one arrives at the
following generic expression\cite{BUV}:
$$\Gamma (H_Q\ra f)=\frac{G_F^2m_Q^5}{192\pi ^3}|KM|^2
\left[ c_3^f\matel{H_Q}{\bar QQ}{H_Q}+
c_5^f\frac{
\matel{H_Q}{\bar Qi\sG Q}{H_Q}}{m_Q^2}+ \right.
$$
$$\left. +\sum _i c_{6,i}^f\frac{\matel{H_Q}
{(\bar Q\Gamma _iq)(\bar q\Gamma _iQ)}{H_Q}}
{m_Q^3} + {\cal O}(1/m_Q^4)\right]  \eqno(4)$$
where the dimensionless coefficients $c_i^f$ depend on the
parton level
characteristics of $f$ (such as the ratios of the final-state quark
masses
to $m_Q$); $KM$ denotes the appropriate combination of KM
parameters,
and $\sG = \sigma _{\mu \nu}G_{\mu \nu}$
with $G_{\mu \nu}$ being the gluonic field strength tensor. The last
term
in eq.(4)
implies also the summation over the four-fermion operators with
different light flavours $q$.
It is through the quantities
$\matel{H_Q}{O_i}{H_Q}$ that the dependence on the {\em decaying
hadron} $H_Q$, and
on
non-perturbative forces in general, enters; they reflect the
fact that the weak decay of the heavy quark $Q$ does not proceed
in empty space, but within a cloud of light degrees of
freedom -- (anti)quarks and gluons -- with which $Q$ and
its decay products can interact strongly.
These are matrix
elements
for on-shell hadrons $H_Q$; $\Gamma (H_Q\ra f)$ is thus
expanded into a power series in $\mu _{had}/m_Q < 1$. For
$m_Q\ra \infty$ the contribution from the lowest dimensional
operator obviously dominates; here it is the dimension-three
operator $\bar QQ$.

There are six important qualitative features
to be noted about this expansion:

\noindent (i) If eq.(1) were indeed correct with the
scale for the transition rate set by the low energy quantity
$\aver{E_{\bar q}}$ rather than by $m_Q$, a $1/m_Q$
expansion would be of dubious, if any, value. Fortunately this
contribution turns out to be spurious for {\em inclusive}
transitions: when all terms, in particular also those coming
from the interference between the WA and the spectator
amplitudes, are summed up, all terms of order
$1/\aver{E_{\bar q}^2}$ and even $1/\aver{E_{\bar q}}$
cancel\cite{MIRAGE};
i.e. an expansion purely in powers of $1/m_Q$
holds for inclusive reactions!

\noindent (ii) Since
$\matel{H_Q}{\bar QQ}{H_Q}=1+
{\cal O}(1/m_Q^2)$, one reads off from eq.(4) that
the leading contribution to the total decay
width is
{\em universal} for all hadrons
of a given heavy-flavour quantum number;
i.e., for $m_Q\ra \infty$ one has
derived -- from QCD proper -- the
spectator picture! This is not a surprising result;
still it is gratifying.

\noindent (iii) Contributions of order $1/m_Q$ would
dominate all other effects -- if they were present! The heavy
quark expansion shows unequivocally that they are absent in
total rates due to a subtle intervention of the local colour gauge
symmetry. This has many important ramifications
\cite{REPORT}: e.g., one can infer that most $B_c$ decays are
driven by the decay of the $\bar c$ inside the $B_c$ meson
and that $\tau (B_c)$ is short, namely well below 1 psec
-- contrary to some claims in the literature.

\noindent (iv) Lifetime differences first arise at order
$1/m_Q^2$ and are controlled by the expectation
values of dimension-five operators. These
terms, which had
been overlooked in the original phenomenological analyses,
generate
a lifetime difference between heavy-flavour {\em baryons} on one
side  and {\em mesons} on the other. Yet apart from
small isospin or $SU(3)_{fl}$ breaking they do {\em not} shift the
meson lifetimes relative to each other.

\noindent (v) Differences in the meson lifetimes emerge at
order $1/m_Q^3$ and are expressed through the expectation values
of four-fermion operators; those are proportional to
$f_M^2$ with $f_M$ denoting the decay constant for the meson
$M$.  Contributions from what is referred to as
WA and PI in the
original phenomenological descriptions are systematically and
consistently included. Further
contributions to the baryon-meson lifetime difference also
arise at this level due to WS.

\noindent (vi) Since the transitions $b\ra c l \nu$ or
$c\ra s l \nu$ are described by an isosinglet operator one can
invoke the isospin invariance of the strong interactions to
deduce for the semileptonic widths
$$ \Gamma _{SL}(B^-)= \Gamma _{SL}(B_d)\; \; \; , \; \; \;
\Gamma _{SL}(D^+)= \Gamma _{SL}(D^0) \eqno(5)$$
and therefore
$$\frac{\tau (B^-)}{\tau (B_d)}=
\frac{BR_{SL}(B^-)}{BR_{SL}(B_d)}\; \; \; , \; \; \;
\frac{\tau (D^+)}{\tau (D^0)}=
\frac{BR_{SL}(D^+)}{BR_{SL}(D^0)} \eqno(6)$$
up to small corrections due to the KM [Cabibbo]
suppressed transition $b\ra u l \nu$
[$c\ra d l \nu$] which changes isospin by half a unit.
The spectator ansatz goes well beyond eq.(5): it
assigns the same semileptonic width to all hadrons of
a given heavy flavour. Yet such a property cannot be
deduced on {\em general} grounds: for one had to rely
on $SU(3)_{Fl}$ symmetry to relate
$\Gamma _{SL}(D_s)$ to $\Gamma _{SL}(D^0)$ or
$\Gamma _{SL}(B_s)$ to $\Gamma _{SL}(B_d)$ and
no symmetry can be invoked to relate the semileptonic
widths of mesons and baryons. There is actually a WA
process that generates semileptonic decays on the
Cabibbo-allowed level for $D_s$ [and also for
$B_c$], but not for the other heavy-flavour states:
the hadrons are produced
by gluon emission off the $\bar s$ [or the $\bar c$] line.
Yet since the relative weight of WA is significantly reduced
in meson decays, one does not expect this mechanism to
change $\Gamma _{SL}(D_s)$ significantly relative to
$\Gamma _{SL}(D^0)$. Contributions to the semileptonic
widths arise already in order $1/m_Q^2$. Yet on rather
general grounds one predicts the expectation values
$\matel{P_Q}{\bar QQ}{P_Q}$ and
$\matel{P_Q}{\bar Qi\sG Q}{P_Q}$ to be largely
independant of the flavour of the light antiquark in the
meson and therefore
$$\Gamma _{SL}(D_s)\simeq \Gamma _{SL}(D^0)\; \; \; , \; \; \;
\Gamma _{SL}(B_s)\simeq \Gamma _{SL}(B_d) \eqno(7)$$
On the other hand,
as explained below, the values of the expectation values
for these operators are different when taken between
baryon states and one expects
$$\Gamma _{SL}(\Lambda _Q) > \Gamma _{SL}(P_Q)
\eqno(8)$$

Through order $1/m_Q^3$ the non-perturbative corrections
in eq.(4) are expressed through the expectation values of three
operators.
A heavy quark expansion yields
$$\matel{H_Q}{\bar QQ}{H_Q} = 1-
\frac{\aver{(\vec p_Q)^2}_{H_Q}}{2m_Q^2}+
\frac{\aver{\mu _G^2}_{H_Q}}{2m_Q^2} + {\cal O}(1/m_Q^3)
\eqno(9)$$
where $\aver{(\vec p_Q)^2}_{H_Q}\equiv
\matel{H_Q}{\bar Q(i\vec D)^2Q}{H_Q}$ denotes the average
kinetic energy of the quark $Q$ moving inside the hadron
$H_Q$ and $\aver{\mu _G^2}_{H_Q}\equiv
\matel{H_Q}{\bar Q\frac{i}{2}\sG Q}{H_Q}$.

\noindent For the chromomagnetic operator one finds
$ \aver{\mu _G^2}_{P_Q}\simeq
\frac{3}{4} (M_{V_Q}^2-M_{P_Q}^2)$,
where $P_Q$ and $V_Q$ denote the pseudoscalar and vector mesons,
respectively.  Therefore
$$\aver{\mu _G^2}_B\simeq 0.37\, GeV \; \;  , \; \;
\aver{\mu _G^2}_D\simeq 0.41\, GeV \eqno(10a)$$
For $\Lambda _Q$ and $\Xi _Q$ baryons one has
instead
$$\aver{\mu _G^2}_{\Lambda _Q, \Xi _Q} \simeq 0 \eqno(10b)$$
since the light diquark system inside $\Lambda _Q$ and $\Xi _Q$
carries no spin.

\noindent For the quantity $\aver{(\vec p_Q)^2}_{H_Q}$
there exists an estimate from a QCD
sum rules analysis\cite{QCDSR} yielding $\aver{(\vec p_b)^2}_B
\simeq 0.5\, \pm \, 0.1\, GeV$ and one can expect one from
lattice QCD in the foreseeable future. We do have a
model-independant lower
bound on it\cite{OPTICAL}, namely $\aver{(\vec p_b)^2}_B \geq
 0.37\, \pm \, 0.1\, GeV$. The
{\em difference} in the kinetic energy of Q inside
baryons and mesons can be related to the masses of charm and
beauty hadrons:
$$ \langle (\vec p_Q)^2\rangle _{\Lambda _Q}-
\langle (\vec p_Q)^2\rangle _{P_Q} \simeq
\frac{2m_bm_c}{m_b-m_c}\cdot
\{ [\langle M_B\rangle -M_{\Lambda _b}]-
[\langle M_D\rangle -M_{\Lambda _c}] \}
\eqno(11)$$
where $\langle M_{B,D}\rangle$ denote the `spin averaged' meson
masses: $ \langle M_B\rangle \equiv \frac{1}{4}(M_B+3M_{B^*})$
and likewise for $\langle M_D\rangle$
\footnote{The crucial assumption here is that $c$ quarks are
sufficiently heavy for a $1/m_c$ expansion to be of
practical help.}. Using data
one finds: $\langle (\vec p_Q)^2\rangle _{\Lambda _Q}-
\langle (\vec p_Q)^2\rangle _{P_Q}= -(0.07 \pm 0.20)(GeV)^2$;
i.e., the present measurement of $M_{\Lambda _b}$
is not yet sufficiently accurate.

\noindent The expectation values for the four-quark operators
taken between {\em meson} states can be expressed in terms of a
single quantity, namely the decay constant:
$$\matel{H_Q(p)}{\bar Q_L \gamma _{\mu}q_L)
(\bar q_L \gamma _{\nu}Q_L)}{H_Q(p)}\simeq
\frac{1}{4} f^2_{H_Q}p_{\mu}p_{\nu} \eqno(12)$$
where factorization has been assumed. The theoretical
expectations for the decay constants are
$$f_D \simeq 200\, \pm 30\, MeV\; \; \; , \; \; \;
f_B \simeq 180 \, \pm 30\, MeV \eqno(13a)$$
$$f_{D_s}/f_D \simeq 1.15 - 1.2\; \; \; , \; \; \;
f_{B_s}/f_B \simeq 1.15 - 1.2 \eqno(13b)$$
The size of the expectation values taken between
{\em baryonic} states are quite uncertain at present. There
exists more than one relevant contraction, and for the time
being quark model estimates provide us with the only guidance!
I will return to this point when discussing predictions of
baryon lifetimes.

While there are significant uncertainties and ambiguities in the
values of the masses of beauty and charm quarks, their difference
which is free of renormalon contributions is
tightly constrained:
$$m_b-m_c\simeq \langle M\rangle _B - \langle M\rangle _D +
\langle (\vec p)^2\rangle \cdot
\left( \frac{1}{2m_c}- \frac{1}{2m_b}\right)
\simeq 3.46 \pm 0.04\, GeV\, .\eqno(14)$$
This value agrees very well with the one extracted from
an analysis of energy spectra in semileptonic $B$ decays
\cite{VOLOSHIN2}.

In summary: (i) One expects on rather general grounds that
the nonperturbative corrections in
{\em beauty} decays amount typically
to no more than a few percent with an expansion parameter
$\sqrt{\aver{\mu _G^2}_B/m_b^2} \sim 0.13$.
(ii) The situation in charm
decays on the other hand is unclear since the expansion parameter
is considerably larger: $\sqrt{\aver{\mu _G^2}_D/m_c^2}
\sim 0.46$.

\pagebreak
{\noindent
\bf IV. Applications and Comparisons with the Data}

\noindent There are three applications I want to discuss here:
semileptonic branching ratios, extracting $V(cb)$ from
$\Gamma _{SL}(B)$ and
the lifetimes of charm and beauty hadrons.

\noindent {\em (A) The semileptonic branching ratio of $B$
mesons}

\noindent The present world average yields\cite{PDG}
$BR_{SL}(B) = 0.1043 \pm 0.0024$. A free parton model treatment
leads to $BR_{SL}(b)|_{PM} \simeq 0.15$ which is lowered by
perturbative QCD corrections:
$BR_{SL}(b)|_{pert.QCD}\simeq 0.125 - 0.135$.
The data differ from this expectation by $\sim 15-20\%$.
A priori one would think that nonperturbative corrections
transforming $BR_{SL}(b)$ into $BR_{SL}(B)$ could naturally
close the gap since they might be of order
$\mu _{had}/m_b \sim 10 - 20\%$ for
$\mu _{had} \sim 0.5 - 1$ GeV. Yet, as stated above, the leading
nonperturbative contributions arise at order
$(\mu _{had}/m_b)^2 \sim 1 - 4\%$. A more detailed analysis
shows that $BR_{SL}(B)$ is indeed lowered relative to
$BR_{SL}(b)$, but only by $\sim 2\%$\cite{BAFFLING}.
There exists a loophole,
though, in that analysis: the energy release in the channel
$b\ra c \bar cs$ is not large and terms in the expansion that
are formally of higher order in $1/m_b$ might actually be quite
significant numerically. There is some theoretical evidence that
they would indeed enhance $\Gamma (B\ra [c\bar cs])$. If
$\Gamma (B\ra [c\bar c s \bar q])\sim 2\cdot
\Gamma (b\ra c \bar cs)$ were to hold, the non-leptonic $B$
width would be enhanced sufficiently to bring the prediction
on $BR_{SL}(B)$ into line with the data.
Such a resolution would have
another observable consequence: it would raise the charm content
$N_c$ of the $B$ decay products quite significantly:
$$N_c|_{expect.} \simeq 1.25 - 1.3 \; \; \; if \; \; \;
\Gamma (B\ra [c\bar c s \bar q])\simeq 2\cdot
\Gamma (b\ra c \bar cs) \eqno(15a)$$
$$N_c|_{expect.} \simeq 1.15 \; \; \; if \; \; \;
\Gamma (B\ra [c\bar c s \bar q])\simeq
\Gamma (b\ra c \bar cs) \eqno(15b)$$
At this meeting we heard about a new preliminary CLEO analysis
yielding\cite{HONSCHEID}
$N_c|_{obs.} = 1.17 \pm 0.05$. This number
-- if true -- would suggest that
the problem of the
`baffling' semileptonic branching ratio is fading away
largely due to a higher than originally observed charm content.
However after this conference I have been informed that the final
CLEO number will be somewhat lower; the issue of
$BR_{SL}(B)$ thus remains unsettled.

Another less publicized puzzle has found its resolution
\cite{BUV}:
the semileptonic branching ratio of $D$ {\em mesons}
through order
$1/m_c^2$ -- i.e., before the $D^+$-$D^0$ lifetime difference
is generated in order $1/m_c^3$ -- is estimated to be around
9\% rather than the
$\sim 15$\% expected for $c$ {\em quarks}. Thus there is
no contradiction with the findings that the
lifetime difference is produced
mainly by PI rather than by WA.

\noindent {\em (B) Extracting $|V(cb)|$ from $\Gamma _{SL}(B)$}

\noindent Measuring $\Gamma _{SL}(B)$ allows to determine
$|V(cb)|$ -- if the total semileptonic  width can reliably be
calculated.
Considerable progress has been achieved over the last few years
in determining the nonperturbative as well as perturbative
corrections. It might
seem at first that these efforts would go for
naught since $\Gamma _{SL}(B)$ depends on the fifth power of the
beauty {\em quark} mass $m_b$ with its intrinsic uncertainties.
However it turns
out that $\Gamma _{SL}(B)$ depends on
$m_b - m_c$ rather than on $m_b$ and $m_c$ separately,
and this difference is
rather tightly constrained, see eq.(14).
With that information one finds\cite{SUV}
$$|V(cb)|_{incl} \simeq (0.0410 \pm 0.002)\cdot
\sqrt{\frac{1.5\, psec}{\tau _B}}\cdot
\sqrt{\frac{BR_{SL}(B)}{0.1043}}\eqno$$
It has been claimed that such an extraction is quite unreliable since
the perturbative expansion of $\Gamma _{SL}(B)$ is ill-behaved:
while the ${\cal
O}(\alpha _S^2)$ corrections have not been fully determined, an
estimate of their weight based on the
BLM-prescription seem to indicate that
they contain large
coeffcients of around 10 - 20!  Closer scrutiny however shows
the following:
if the theoretically sound
`running' mass evaluated around a scale of 1 GeV is employed,
the expansion is well-behaved; i.e., the large corrections get absorbed
into the definition of the quark mass\cite{SU}.

\noindent {\em (C) Charm Lifetimes}

The expectations \cite{MIRAGE,DS}
for the lifetimes of charm hadrons through order
$1/m_c^3$ are juxtaposed to  the data \cite{MALVEZZI} in
Table \ref{TABLE1}.
\begin{table}
\begin{tabular} {|l|l|l|}
\hline
Observable &QCD ($1/m_c$ expansion) &Data \\
\hline
\hline
$\tau (D^+)/\tau (D^0)$ & $\sim 2$
 [for $f_D \simeq$ 200 MeV]
&$2.547 \pm 0.043$ \\
\hline
$\tau (D_s)/\tau (D^0)$ &$1\pm$ few \%
&  $ 1.125\pm 0.042$ \\
\hline
$\tau (\Lambda _c)/\tau (D^0)$&$\sim 0.5\; ^*$ &
$0.51\pm 0.05$\\
\hline
$\tau (\Xi ^+ _c)/\tau (\Lambda _c)$&$\sim 1.3\; ^*$ &
$1.75\pm 0.36$\\
\hline
$\tau (\Xi ^+ _c)/\tau (\Xi ^0 _c)$&$\sim 2.8\; ^*$ &
$3.57\pm 0.91$\\
\hline
$\tau (\Xi ^+ _c)/\tau (\Omega _c)$&$\sim 4\; ^* $&
$3.9 \pm 1.7$\\
\hline
\end{tabular}
\centering
\caption{QCD Predictions for Charm Lifetimes}
\label{TABLE1}
\end{table}
The agreement with the data is remarkable
considering that the expansion
parameter is not much smaller than unity here.
A few more detailed comments are
in order:

\noindent $\bullet$ The $D^+$-$D^0$ lifetime difference is driven
mainly by PI with WA contributing not more than 10 - 20\%.
Including renormalization down to $\mu _{had}$ is
numerically essential. Within the accuracy of the expansion, the
data are reproduced.

\noindent $\bullet$ The $D_s$-$D^0$ lifetime ratio can be treated
with better theoretical accuracy, namely of order a few percent.
The observed near equality of $\tau (D_s)$ and $\tau (D^0)$
represents rather direct evidence for the
{\em reduced} weight of WA in
charm {\em meson} decays\cite{DS}.

\noindent $\bullet$ The success so far in predicting
baryon lifetime ratios is even more remarkable, since the
baryonic widths receive contributions of {\em both signs}
and the
relevant expectation values are computed in quark models,
as indicated by the asterisk in the Table.
\footnote{In passing one should note that a new element enters
in $\Gamma (\Omega _c)$: $\aver{\mu _G^2}_{\Omega _c}
\neq \aver{\mu _G^2}_{\Lambda _c} \simeq 0$ since the light
di-quark system inside $\Omega _c$ carries spin one.}

\noindent {\em (D) Beauty Lifetimes}

Quantitative predictions \cite{STONE2}
for the lifetime ratios of beauty
hadrons through order $1/m_b^3$ are given in Table \ref{TABLE2}
together with present data \cite{SHARMA}.
\begin{table}
\begin{tabular} {|l|l|l|}
\hline
Observable &QCD ($1/m_b$ expansion) &Data \\
\hline
\hline
$\tau (B^-)/\tau (B_d)$ & $1+
0.05(f_B/200\, \MeV )^2
[1\pm {\cal O}(10\%)]>1$ &$1.04 \pm 0.05$ \\
&(mainly due to {\em destructive} interference) & \\
\hline
$\bar \tau (B_s)/\tau (B_d)$ &$1\pm {\cal O}(0.01)$
&  $ 0.98\pm 0.08$ \\
\hline
$\tau (\Lambda _b)/\tau (B_d)$&$\sim 0.9\; ^*$ & $0.76\pm 0.06$
\\
\hline
\end{tabular}
\centering
\caption{QCD Predictions for Beauty Lifetimes}
\label{TABLE2}
\end{table}
The predictions follow the same general pattern as in charm
decays. Yet the deviations of the lifetime ratios from unity
are much smaller since $1/m_b^2 \ll 1/m_c^2$; for the same
reason one has more faith in the reliability of the $1/m_Q$
expansion for beauty than for charm decays.

The near-equality with $\tau (B_d)$ refers to the {\em average}
$B_s$ lifetime, $\bar \tau (B_s)$. For
the difference in the lifetimes of the two
$B_s$ mass eigenstates one predicts\cite{BSBS}
$$\frac{\Delta \Gamma (B_s)}{\bar \Gamma (B_s)}
\equiv \frac{\Gamma (B_{s,short})-\Gamma (B_{s,long})}
{\bar \Gamma (B_s)}\simeq 0.18\cdot
\frac{(f_{B_s})^2}{(200 \, MeV)^2}\, ,  \eqno(17)$$
i.e., the largest lifetime difference among $B^-$,
$B_d$ and $B_s$ mesons is expected to be generated by
$B_s-\bar B_s$ oscillations!
One can search for the existence of two different $B_s$ lifetimes by
comparing $\tau (B_s)$ as measured in
$B_s \ra \psi \eta /\psi \phi$ on one hand and in
$B_s\ra l \nu X$ on the other:
$$\tau (B_s \ra l \nu D^{(*)}) -
\tau (B_s\ra \psi \eta /\psi \phi ) \simeq
\frac{1}{2} [\tau (B_{s,long})-\tau (B_{s,short})]\eqno(18) $$
Whether an effect of the size predicted in eq.(17)
is large enough to be ever observed
in a real experiment, is unclear. Nevertheless one should search
for it even if one has sensitivity only for a 50\% lifetime difference.
For while eq.(17) represents the best presently available
estimate, it is not a `gold-plated' prediction.
It is conceivable that the
underlying computation {\em underestimates} the actual lifetime
difference!

The prediction on $\tau (\Lambda _b)/\tau (B_d)$ appears to be in
serious (though not yet conclusive) disagreement with the data. The
details of what went into that prediction can be found in
ref.\cite{REPORT}; here I want to state only the following conclusion.
If
$\tau (B_d)$ indeed exceeds $\tau (\Lambda _b)$ by 25 - 30 \%,
then a `theoretical price' has to be paid:
(i) The charm mass
represents too low of a scale for allowing to go beyond merely
qualitative
predictions on charm baryon (or even meson) lifetimes,
since it appears that corrections of order $1/m_c^4$ and higher
are still important;  (ii) the present agreement between theoretical
expectations and data on charm baryon lifetimes is largely
accidental and most likely would not survive in the face of more
precise measurements! At the same time an intriguing
puzzle arises: Why are the quark model results for the
relevant expectation values so much off the mark for
beauty baryons? It is the deviation from unity in the lifetime
ratios that is controlled by these matrix elements; finding
a 30 \% difference rather than the expected 10 \% then repreents
a 300 \% error!

\noindent {\em (E) The Ratios of Semileptonic Branching Ratios}

\noindent Both $\Gamma _{SL}(\Lambda _c)$ and
$\Gamma _{NL}(\Lambda _c)$ are predicted to differ
substantially from the corresponding quantities for $D^0$
mesons. There is no intrinsic reason
why $BR_{SL}(\Lambda _c)\simeq
BR_{SL}(D^0)\times \tau (\Lambda _c)/\tau (D^0)
\simeq 0.5\, BR_{SL}(D^0)$ should hold; the
semileptonic $\Lambda _c$ branching ratio is probably
larger than that.
Analogous considerations  lead to
$$\aver{BR_{SL}(beauty)}< \aver{BR_{SL}(B)}\, ,
\eqno(19)$$
where $\aver{BR_{SL}(beauty)}$ denotes the average over
{\em all} beauty hadrons and $\aver{BR_{SL}(B)}$ that
over $B$ mesons.

\noindent{\bf V. Summary and Outlook}

\noindent
{\em Inclusive} heavy-flavour decays can be treated through
an expansion in $1/m_Q$ which allows to express the
leading nonperturbative
corrections through the expectation values of a small number of
dimension-five and -six operators. Basically all such matrix
elements relevant for {\em meson} decays can reliably be related to
other
observables; this allows to extract their size in a
model-independant way.  For {\em baryon} decays, however, one
has at present to rely on quark model calculations to determine
the expectation values of the dimension-six operators relevant for
lifetime differences. The numerical results of such computations
are of dubious reliability; predictions for lifetime ratios
involving heavy-flavour {\em baryons} therefore suffer from larger
uncertainties than those involving only mesons.

In addition to providing us with a more satisfying theoretical
framework the $1/m_Q$ expansion yields also practical benefits:
it reproduces the charm lifetime ratios within the expected
(rather sizeable) uncertainties due to higher order terms; it
predicts unequivocally small differences among
$\tau (B^-)$, $\tau (B_d)$ and $\bar \tau (B_s)$.

At present there exists one glaring phenomenological
problem and -- not surprisingly, as
just indicated -- it concerns baryon decays: the observed
$\Lambda _b$ lifetime is
shorter than predicted relative to the $B_d$ lifetime.
Unless future measurements move it up significantly, one
has to pay a theoretical price for that failure.
To the degree that the observed value for
$\tau (\Lambda _b)/\tau (B_d)$ falls below 0.9 one has to draw the
following conclusion: one cannot trust
the numerical results of quark model
calculations for {\em baryonic} matrix elements -- not very
surprising by itself; yet furthermore and more seriously
it would mean that $1/m_Q^4$ or even higher order
contributions are still relevant in charm baryon decays before
fading away for beauty decays.
Then one had
to view the apparently successful predictions on the lifetimes
of charm baryons as largely coincidental!

The theoretical analysis of the lifetimes of heavy-flavour hadrons
can be improved, refined and extended:
(a) {\em improved} by a better
understanding of
quark-hadron duality\cite{MISHA,OPTICAL,REPORT};
(b) {\em refined} by a reliable determination of
in particular, but not only, the baryonic expectation values of
the relevant dimension-six operators;
(c) {\em extended} by treating $\Xi _b$ decays.

There is a host of future measurements that will probe and advance
our understanding of heavy flavour decays:

\noindent (i) While there is no theoretical need to measure
$\tau (D^+)/\tau (D^0)$ [or $\tau (\Lambda _c)/\tau (D^+)$]
more precisely, it is desirable to determine
$\tau (D_s)/\tau (D^0)$ with an accuracy of $\sim$ 1\%;
this will allow us to quantitatively address some aspects
of WA that, despite their subtle nature, are important not
only for $\Gamma (D_s)$ decays, but also for lepton spectra
in $D_s$ and $B^-$ decays \cite{DS}.

\noindent (ii) Measurements of $\Xi ^{0,+}$ and
$\Omega _c$ lifetimes with at least a 10\%
accuracy are clearly needed. Then one could extract
the weight of the various contributions quantitatively
and compare it with their theoretical evaluation; this
in turn would enable us to isolate discrepancies
between data and predictions in charm decays and
shed light on problems we encounter when extrapolating
to beauty baryon decays.

\noindent (iii) Likewise one has to measure
$\tau (\Lambda _c)$, $\tau (\Xi _c^-)$ and $\tau (\Xi _c^0)$
{\em separately}.

\noindent (iv) It is mandatory to confirm that
$\tau (B_d)\simeq \bar \tau (B_s)$ holds within an
accuracy of very few percent
and to verify that $\tau (B^+)$ {\em exceeds}
$\tau (B_d)$ by a few to several percent.
A {\em future} discrepancy between the predictions on
$\tau (B^+)/\tau (B_d)$ or
$\tau (B_d)/\bar \tau (B_s)$ and the data  -- in particular an
observation that the lifetime for $B^+$ mesons is
{\em shorter}
than for $B_d$ mesons -- would have quite fundamental
consequences. For the leading deviation of these ratios from
unity arises at order $1/m_b^3$ and should provide a good
approximation since the expansion parameter is small:
$\mu _{had}/m_b \sim 0.13$. The size of this term is given by
the expectation value of a four-fermion operator expressed in terms
of $f_B$. A failure in this simple situation would
raise very serious doubts about the validity or at least
the practical relevance
of the $1/m_Q$ expansion for treating fully
inclusive nonleptonic transitions;
at best this would leave
semileptonic transitions in the domain of their applicability.
Such a breakdown of quark-hadron duality would
{\em a priori} appear
as a quite conceivable and merely disappointing
scenario. However such an
outcome would have to be seen as quite puzzling
{\em a posteriori}; for in our
analysis we have not discerned any sign indicating the existence
of such a fundamental problem or a qualitative distinction between
nonleptonic and semileptonic decays \cite{BLOKMANNEL,OPTICAL}.
Thus even a
failure would
teach us a valuable, albeit sad lesson about the intricacies of the
strong interactions;  for the heavy quark expansion is
directly and unequivocally based on QCD with the only additional
assumption concerning the workings of
quark-hadron duality!

{\bf Acknowledgements:} I am deeply indebted to
my collaborators for their unusual generosity with which they
shared their insights with me. It has been a genuine pleasure
to attend this inspiring and beautifully organized meeting
in Pisa which houses
one of the most glorious ensemble of Western architecture.
Special treats were provided for all the senses of the
participants, in particular also for their ears! This work was
supported by the National Science Foundation under
grant number PHY 92-13313.  I also thank the Institute for
Nuclear Theory at the University of Washington for its
hospitality and the Department of Energy for partial
support during the completion of this manuscript.

\end{document}